\documentclass[epj]{webofc}
\usepackage[varg]{txfonts}   
\woctitle{International Conference on New Frontiers in Physics 2013}
\newcommand{\dau}{\mbox{$d$$+$Au}\xspace}
\newcommand{\pt}{\mbox{$p_T$}\xspace}
\newcommand{\auau}{\mbox{Au$+$Au}\xspace}
\newcommand{\cucu}{\mbox{Cu$+$Cu}\xspace}
\newcommand{\rda}{\mbox{$R_{dA}$}\xspace}
\newcommand{\ncoll}{\mbox{$N_{\rm coll}$}\xspace}
\newcommand{\jpsi}{\mbox{$J/\psi$}\xspace}
\newcommand{\sqsntwo}{\mbox{$\sqrt{s_{_{NN}}}=200$~GeV}\xspace}
\newcommand{\pau}{\mbox{$p$$+$Au}\xspace}
\newcommand{\pp}{\mbox{$p$$+$$p$}\xspace}
\begin{document}
\title{Open heavy flavor measurements in \dau collisions at PHENIX experiment}
%
%

\author{Sanghoon Lim\inst{1}\fnsep\thanks{\email{shlim@bnl.gov}} for the PHENIX collaboration}

\institute{Physics Department, Yonsei University, Seoul 120-749, Korea}

\abstract{
The heavy quarks produced in the early stage of heavy-ion collisions are very effective probes of the dense partonic medium produced at RHIC. 
PHENIX has the ability to measure heavy quark production through single electrons in the central arm spectrometers ($|\eta|<0.35$) and single muons in the forward (backward) muon spectrometers ($1.2<|\eta|<2.2$). 
As these single leptons are from open heavy-flavor meson semi-leptonic decays, initial state cold nuclear matter effects on heavy quark production can be probed by measuring the single leptons in \dau collisions. 
PHENIX have observed a large enhancement of heavy-flavor electrons in \dau collisions at mid-rapidity, which indicates strong CNM effects on heavy quark production, in contrast to the suppression observed in \auau collisions.
Measurement of single muons from open heavy flavor in \dau collisions at forward (backward) rapidity provide detailed look into rapidity dependent CNM effects as well as the low (high) $x$ parton distribution function within Au nucleus.
We discuss recent PHENIX heavy flavor measurements and how they expand our understanding of CNM effects and contribute to the interpretation of other results in heavy-ion collisions.
}
\maketitle
\section{Introduction}
\label{intro}
Heavy quarks, mostly charm and bottom, are produced in the early stage of heavy-ion collisions, therefore they are good probes to study the evolution of hot and dense medium which is expected to be produced in the heavy-ion collisions. 
Heavy quark production has been studied in various collision systems. 
In case of \pp collision, we can test our theoretical predictions based on perturbative QCD (pQCD) for heavy quark production. 
In \dau and heavy-ion collisions, we can study initial- and final-state modifications by comparing with the results from \pp collisions.

The PHENIX experiment has excellent capabilities to measure leptons from semi-leptonic decay of heavy-flavor mesons, $D$ and $B$. 
In central arms ($|\eta|<0.35$) and muon arms ($1.2<|\eta|<2.2$), electrons and muons from open heavy-flavor, respectively, can be measured by using hadrorn cocktail method. 
At mid-rapidity, a huge suppression of heavy-flavor electron production relative to the scaled \pp results are observed in the central \auau collisions~\cite{ppg066}, whereas a clear enhancement is seen in the central \dau collisions~\cite{ppg131}. 
These results indicate that the suppression of heavy quark production in \auau collisions is due to the hot nuclear matter effects. 
In addtion, the enhancement in central \dau collisions suggests strong cold nuclear matter (CNM) effects. 
At forward rapidity, a similar level of suppression, seen in the central \auau collisions at mid-rapidity, is observed in the central \cucu collisions~\cite{ppg117}, and the pQCD prediction considering additional CNM effects well describe the large suppression at forward region. 

\section{Cold Nuclear Matter Effects}
\label{cnm}
In heavy-ion collisions, the CNM effects are convoluted in the effects from hot and dense medium, so that it is hard to interpret the results in heavy-ion collisions solely with the hot nuclear effects. In order to study the CNM effects, \dau collisions are used as a control experiment. 
There are several CNM matter effects such as modification of nuclear parton distribution function (nPDF), \pt broadening, initial-state energy loss, and nuclear breakup. 
The \rda of heavy-flavor electrons at mid-rapidity suggests a mainly contributed CNM effect at this region is the \pt broadening.
Precise measurements of heavy quark production at other rapidity ranges will help to the detailed study of CNM effects.

\begin{figure}
\centering
\includegraphics[width=0.49\textwidth,clip]{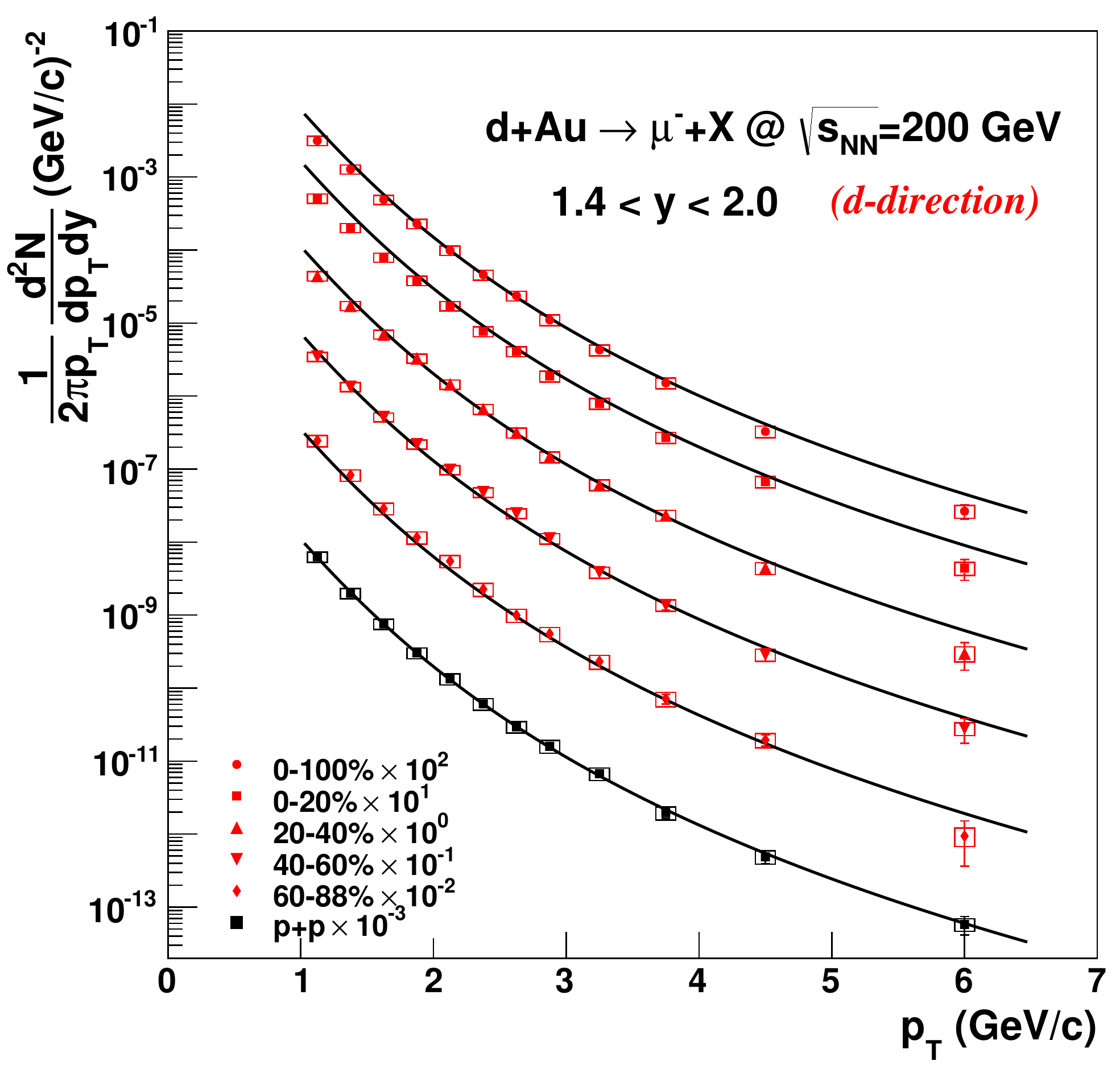}
\includegraphics[width=0.49\textwidth,clip]{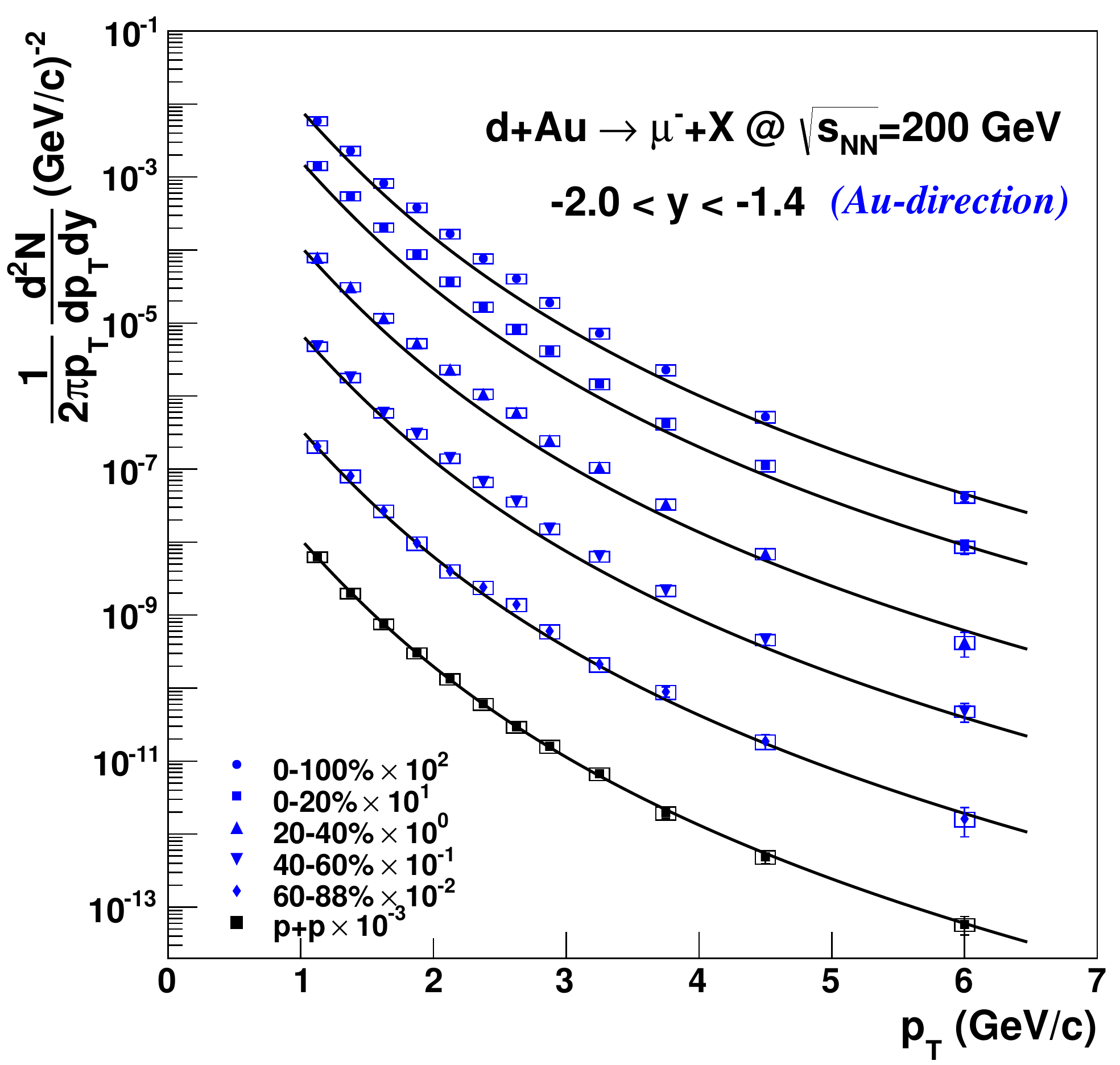}
\caption{Invariant yield of heavy-flavor muons as a function of \pt in \pp and different centrality of \dau collisions at \sqsntwo at forward (left) and backward (right) rapidity. Solid lines are a fit to the \pp results, scaled by the corresponding \ncoll for centrality classes.}
\label{fig0}
\end{figure}

Recently, negatively charged muons from open heavy-flavor decay have been measured in \dau and \pp collisions~\cite{ppg153}.
Figure~\ref{fig0} shows the measured \pt of heavy-flavor muons in different centrality bins of \dau collisions at forward ($1.4<y<2.0$, $d$-going direction) and backward ($-2.0<y<-1.4$, Au-going direction) rapidity regions.
The black square data points at the bottom of both panels represent the \pt spectrum in \pp collisions. 
The new \pp results are consistent with the previous PHENIX measurements~\cite{ppg117} within systematic uncertainty, as well as the systematic uncertainty is reduced based on larger statstics of data and improved analysis techniques.

\begin{figure}
\centering
\includegraphics[width=0.49\textwidth,clip]{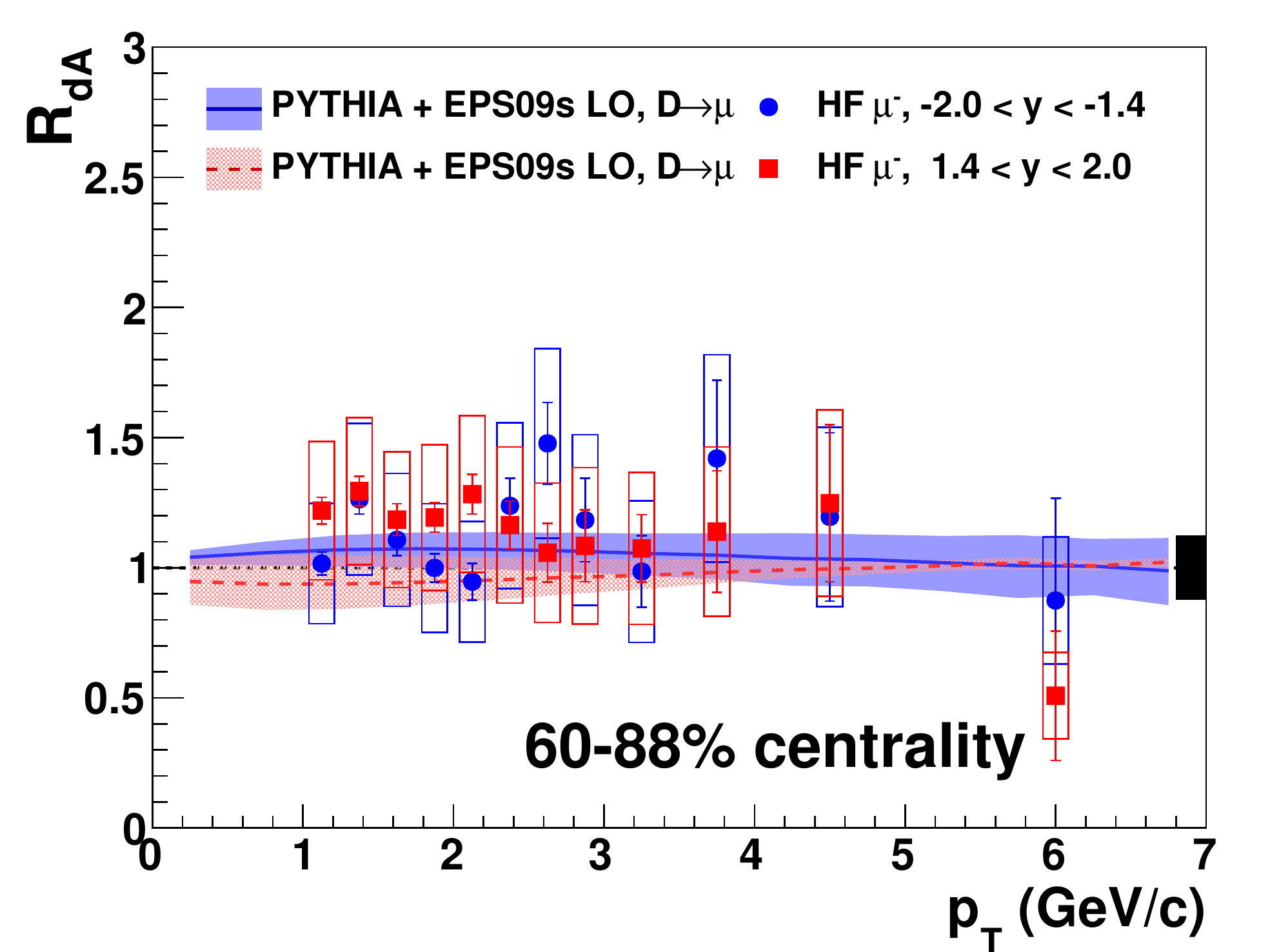}
\includegraphics[width=0.49\textwidth,clip]{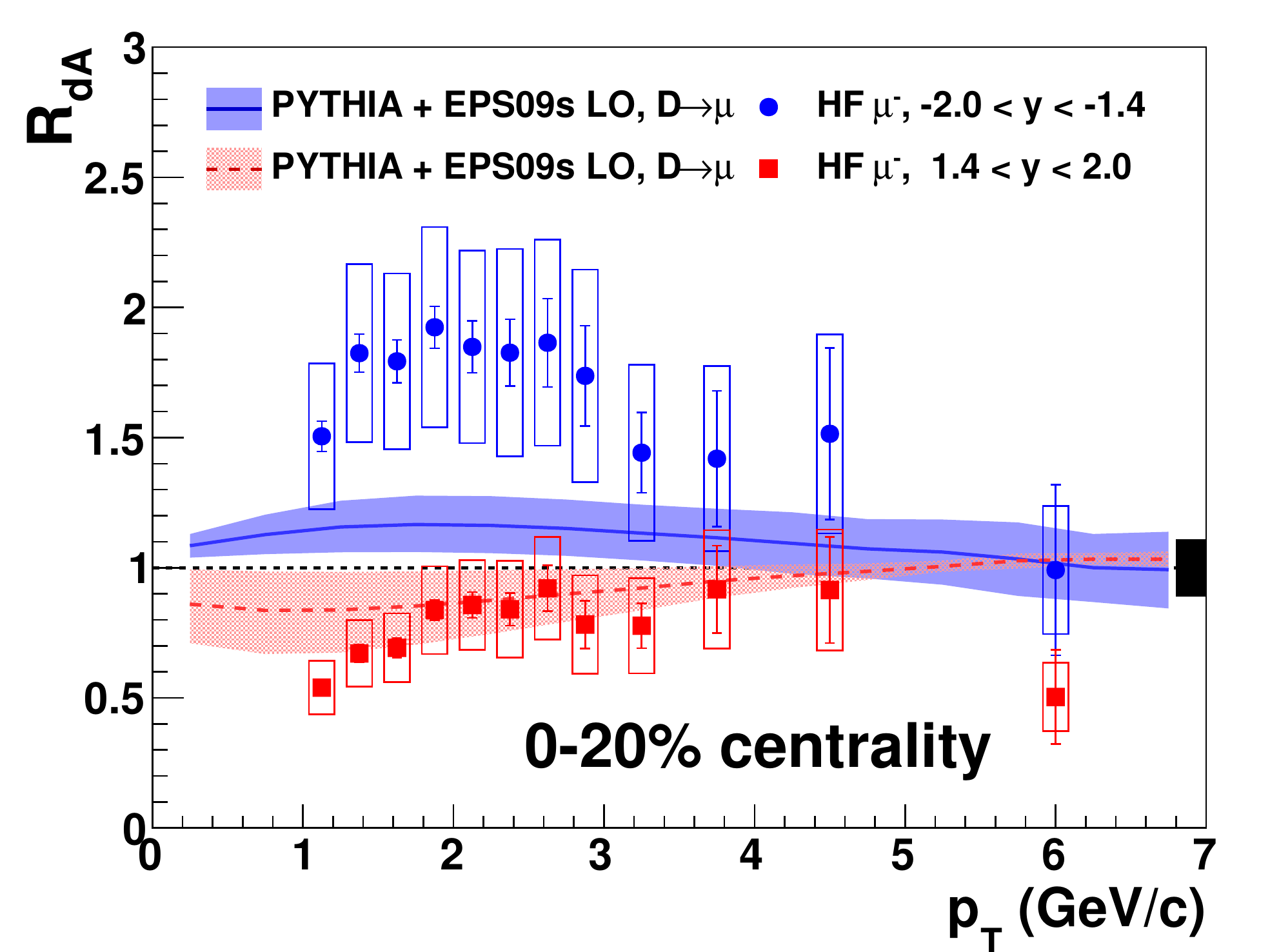}
\includegraphics[width=0.49\textwidth,clip]{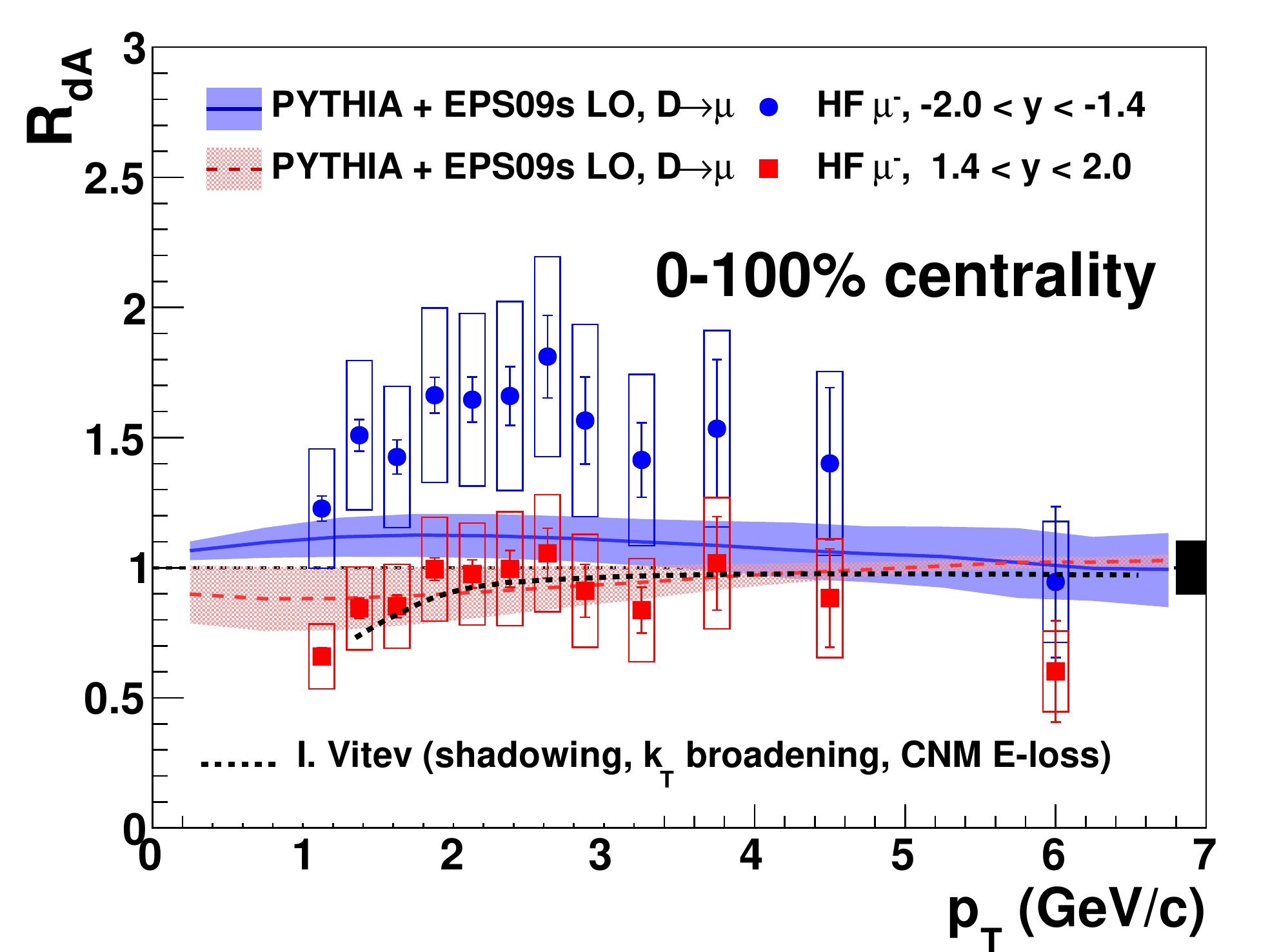}
\caption{The nuclear modification factor \rda as a function of \pt for heavy-flavor muons at forward (red squares) and backward (blue circles) rapidity ranges in the most peripheral (top left), the most central (top right), and the unbiased (bottom) \dau collisions. The red dashed (blue solid) lines in each panel are the  PYTHIA$+$EPS09s nPDF calculations at forward (backward) rapidity. The black dotted line is a pQCD prediction considering the CNM effects for forward rapidity.}
\label{fig1}
\end{figure}

Figure~\ref{fig1} shows the nuclear modification factor \rda as a function \pt for the three centrality classes, 60--88\% (top left), 0--20\% (top right), and 0--100\% (bottom) at forward (red squares) and backward (blue circles) rapidity.
In the most peripheral collisions, the \rda at both rapidity regions are consistent with each other and the unity, indicating no overall modification.
However, strong CNM effects on heavy quark production are observed in the central \dau collisions.
A clear enhancement of heavy-flavor muon production relative to the scaled \pp results is observed at backward rapidity, whereas a suppression is seen at forward rapidity region.
Two bands in each panel are PYTHIA 8 calculations of the $D\to\mu$ process considering nPDF modification based on the EPS09s nPDF set~\cite{eps09s}.  
This theoretical predictions qualitively describe the forward data but underestimate the enhancement seen in the most central collisions.
Another theoretical prediction from pQCD calculation considering three CNM effects, such as shadowing, \pt broadening, and energy loss also shows a good agreement with the forward data.
Therefore, other CNM effects beyond the nPDF modification may significantly contribute to heavy quark production at forward and backward rapidity regions.
 
\begin{figure}
\centering
\includegraphics[width=0.49\textwidth,clip]{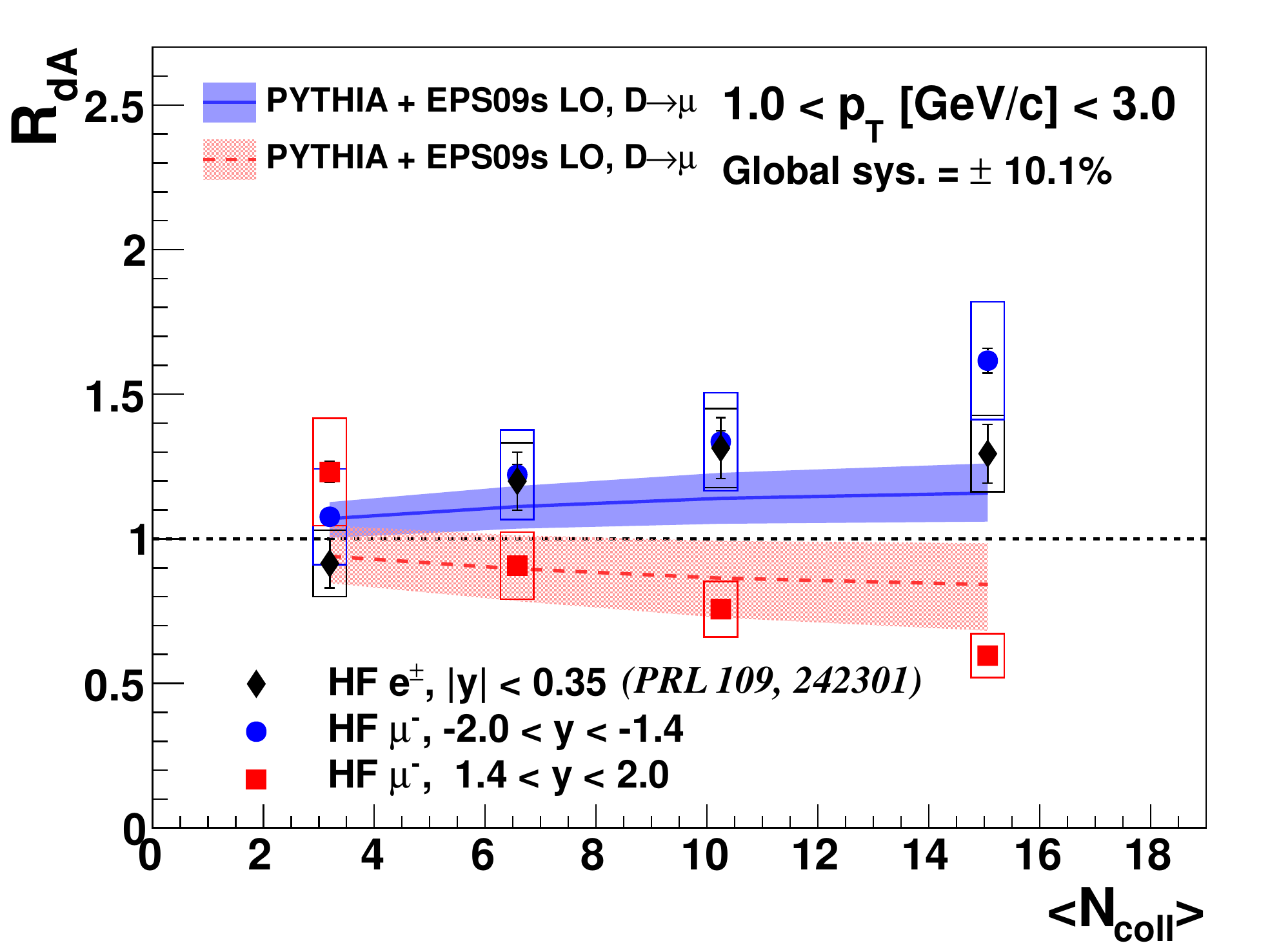}
\includegraphics[width=0.49\textwidth,clip]{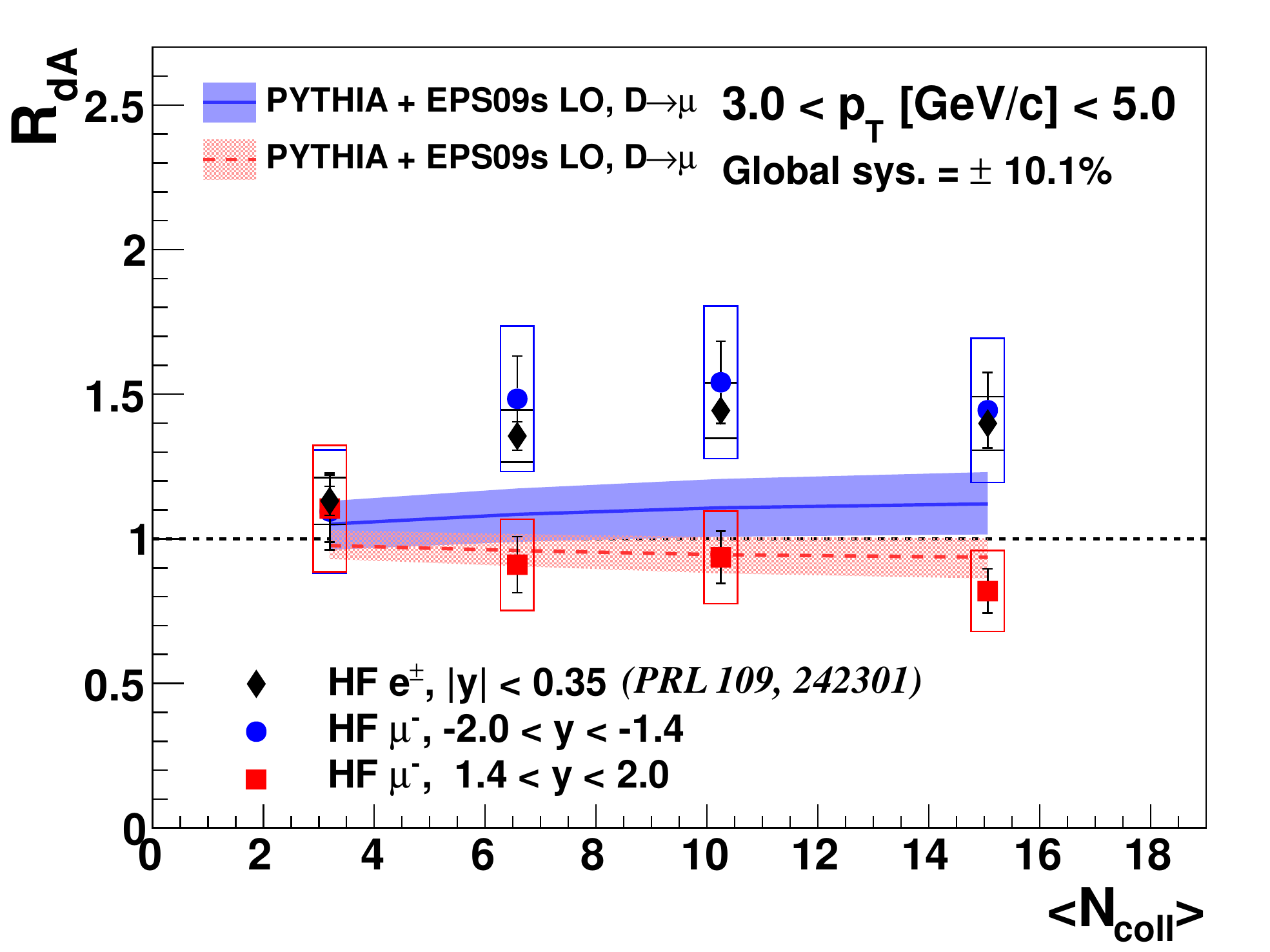}
\caption{The nuclear modification factor \rda as a function of \ncoll for heavy-flavor leptons from different rapidity and \pt bins. The lines with bands are the PYTHIA$+$EPS09s nPDF calculations for forward (red dashed) and backward (blue solid) rapidity regions.}
\label{fig2}
\end{figure}

Figure~\ref{fig2} shows the nuclear modification factor \rda as a function of \ncoll for two different \pt ranges, $1<\pt<3~{\rm GeV}/c$ (left) and $3<\pt<5~{\rm GeV}/c$ (right).
In both \pt ranges, the suppression (enhancement) becomes larger with increasing centrality.
The results at mid-rapidity (black circles) shows a similar trend with that seen in the backward data.
The EPS09s nPDF calculations does not reproduce the large difference between forward and backward rapidity.

\begin{figure}
\centering
\includegraphics[width=0.6\textwidth,clip]{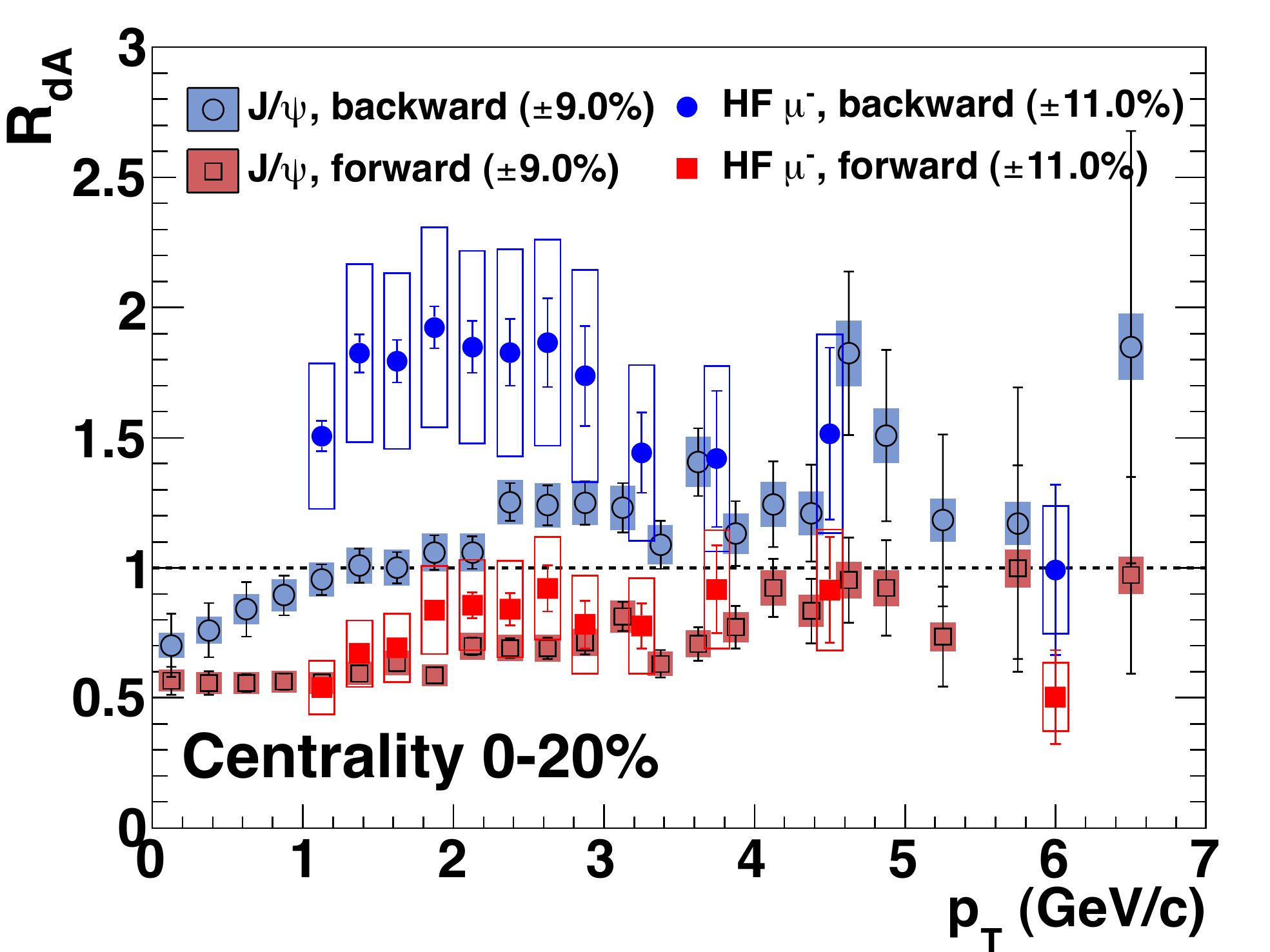}
\caption{Comparison of the nuclear modification factor \rda between \jpsi and heavy-flavor muons in the most central \dau collisions.}
\label{fig3}
\end{figure}

Quakonia is additionally affected by the nuclear breakup, interaction with surrounding matters.
Figure~\ref{fig3} shows comparisons of \rda as a function of \pt between \jpsi~\cite{ppg125} and heavy-flavor muons in the most central \dau collisions at forward and backward rapidity.
At forward rapidity, \rda of \jpsi and heavy-flavor muons are consistent within the uncertainties.
However, the backward results show a large difference at $\pt<3~{\rm GeV/c}$, suggesting a significant breakup effect at backward rapidity.
These data may provide a new contraint on theoretical models for quarkonia production.

\section{Summary}
\label{summary}
Negatively charged muons from heavy-flavor meson decay in various centrality classes of \dau collisions at \sqsntwo have been measured at forward and backward rapidity ranges. For the most peripheral centrality class, The \rda at both rapidity are consistent with each other and the unity indicating no overall modification. In the most central collisions, an clear enhancement of heavy-flavor muon production is observed at backward rapidity, whereas a suppression is seen at forward rapidity region. The pQCD prediction considering the CNM effects shows a good agreement with the forward data. In addition, another theoretical calculation based on the EPS09s nPDF set qualitatively reproduce the forward \rda as well. However this calculation, considering only the modification of nPDF, underestimate the difference between forward and backward rapidity regions observed in the data, suggesting the significant role of other CNM effects beyond the modification of parton density functions. 

The \rda as a function of \ncoll for two different \pt bins show a larger enhancement (suppression) as increasing centrality at backward (forward) rapidity. The trend seen in backward rapidity is similar with the heavy-flavor electron results at mid-rapidity. The comparison to the \jpsi results suggets a significant role of the nuclear breakup in the quarkonia production.

New silicon vertex detectors (VTX/FVTX) was installed, and these systems will provide high precision of primary and secondary vertex measurements. The  enhanced information will allow to measure charm ($D$) and bottom ($B$) production separately. These new measurements with the data in upcoming runs (\auau--RHIC Run-14, \pau--RHIC Run-15) will help to improve the current understanding of hot and cold nuclear matters as well as provide essential constraints on theoretical predictions.


\begin{thebibliography}{}
\bibitem{ppg066}
Adare, A. and others (PHENIX collaboration), Phys. Rev. Lett. \textbf{98}, 172301 (2007)
\bibitem{ppg131}
Adare, A. and others (PHENIX collaboration), Phys. Rev. Lett. \textbf{109}, 242301 (2012)
\bibitem{ppg117}
Adare, A. and others (PHENIX collaboration), Phys. Rev. \textbf{C86}, 024909 (2012)
\bibitem{ppg153}
Adare, A. and others (PHENIX collaboration), arXiv:1310.1005 (2013)
\bibitem{eps09s}
Helenius, I. and Eskola, K. and Honkanen, H and Salgado, C JHEP \textbf{1207} 073 (2012)
\bibitem{ppg125}
Adare, A. and others (PHENIX collaboration), Phys. Rev. \textbf{C87}, 034904 (2013)

%
%
\end{thebibliography}
\end{document}